Sand and fire: applying the sandpile model of self-organized criticality to wildfire mitigation

Running head: Sandpile model: prescribed fire & reduced wildfire

Joshua E. Gang[1+] Wanqi Jia[2+], and Ira A. Herniter[3]*

[1]Google LLC, Mountain View, CA 94043
[2]University High School, Irvine, CA 92612
[3]Department of Plant Biology, Rutgers University, New Brunswick, NJ 08901
[+]These authors contributed equally to this work
*Corresponding author: iah17@sebs.rutgers.edu; 59 Dudley Road, Room 379, New Brunswick, NJ 08901; +1-848-229-4255

Abstract

Prescribed burns have been increasingly utilized in forest management in the past few decades. However, their effectiveness in reducing the risk of destructive wildfires has been debated. The sandpile model of self-organized criticality, first proposed to model natural hazards, has been recently applied to wildfire research for describing a negative linear relationship between the logarithm of fire size, in area burned, and the logarithm of the fire incidence number of that size. In this study, we leverage this powerful sandpile model to perform a series of simulations demonstrating that prescribed fires indeed suppress the incidence of destructive wildfires. The same conclusion was obtained when this method was utilized to analyze historical data of forest fires from three American states: Florida, California and Georgia. Our study justifies the application of the sandpile model to wildfire research and establishes a novel method based on the analysis of 'slope' estimated from the sandpile model, related to the 'leverage' described in the literature, for facilitating the investigation of potential risk factors of wildfires. The new method may be utilized for the development of optimal strategies for prescribed burning. An R-script for sandpile model simulation is available for further wildfire investigation.

Brief summary

Both simulation and analysis of historical forest fire data leveraging the sandpile model suggest that prescribed burning generally reduces the risk of destructive wildfires.

Introduction

Wildfire is of increasing risk to people and landscapes as anthropogenic climate change progresses (IPCC 2018). The 2018 Camp Fire in Northern California was the largest and most destructive fire in California history, burning 153,336 acres (620.5 km$^2$) and responsible for eighty-five fatalities (Cal Fire 2019). Even if global warming can be limited to just 1.5°C above preindustrial levels, the incidence and intensity of wildfires is expected to increase dramatically (Settele *et al*. 2015).

While wildfires can be deadly and destructive, fire also plays a major role in promoting ecological diversity and maintaining sustainability in bio-systems. For example, regular fires result in increased diversity of bat species (Steel *et al*. 2019), and some plant species require fire treatment of their seeds for germination (Keeley 1987). Efforts to prevent as many fires as possible over the past century by federal and state agencies in the United States have resulted in unhealthy buildups of brush and dead wood littering forest floors, promoting more intense and destructive fires (Haugo *et al*. 2019; Roos *et al*. 2020; Minnich and Chou 1997). Prescribed fires, as defined by the National Park Service (2020), are intentionally set, low intensity fires, which are used to manage land. The main hypothetical function of prescribed burning is to reduce the ground litter, which serves as fuel driving the most destructive wildfires. There is evidence to show that regularly prescribed burning of forests may be highly beneficial by, among other effects, maintaining critical species habitat (Russell *et al*. 1999). Other research, however, has claimed that the association between previous fires and reduction of the area of wildfire does not generally hold in global case studies from a range of biomes (Bradstock *et al*. 2012; Price *et al*. 2012; Price *et al*. 2015). The hypothesis that prescribed burns can significantly reduce the risk of destructive wildfires needs to be further tested using properly selected models, in which other covariates need to be well controlled.

The sandpile model of self-organized criticality was introduced by Bak *et al.* (1987). In this model, the main ingredients include (1) slow driving towards some instability and (2) a mechanism to relax tension locally and partially. The model consists of an imagined plane organized as a grid, where a large number of individual grains of sand are dropped onto the plane in a random fashion. In each site of the grid, grains of sand can be placed. Fig. 1 contains an example of such a grid. Once the number of grains reaches a predefined threshold, the pile in the site collapses, transferring its collected grains to neighboring sites in the cardinal directions. If any of those sites then reach the threshold, the piles in the adjacent sites collapse and transfer their collected grains to their respective neighboring sites. When placing a sand grain causes a pile collapse on a site, this collapse can propagate across the plane until the plane reaches equilibrium without affecting sites on the borders of the plane, or until the propagating collapse impacts on the border(s) to cause an avalanche – an event of interest in the process. The size of an event (avalanche) is represented by the number of hit(s) on the boundaries, i.e., a propagating collapse that reaches the borders of the plane. Testing this model has shown that the size of avalanches and the frequency of avalanches of that size follows a power-law, which takes the form of

$$f(A_c) = \alpha A_c^{-\beta}, \tag{1}$$

where $A_c$ is the size of collapse (the number of hit(s) on the plane boundaries) and the $f(A_c)$ is the corresponding frequency (Bak *et al*. 1987; Cajueiro and Andrade 2010). In examining this effect, it has been found that numerous real-world phenomena follow this same pattern, including earthquake power (Bak and Tang 1989), solar x-ray flares (Crosby *et al*. 1993), and even evolutionary rates (Sneppen *et al*. 1995). Of particular relevance, the sandpile model has been successfully applied to natural wildfire systems to demonstrate the relationship between wildfire sizes and their frequencies (Ricotta *et al.* 1999; Malamud 2004; Yoder *et al.* 2011). The analysis of historical wildfire data from "natural experiments" using the sandpile model showed that small early season fires, left to burn to natural extinction, contribute to a region's natural fire resistance (Yoder *et al.* 2011). This conclusion suggests that prescribed burning likely provides additional resistance.

Ishii *et al*. (2002) showed that agent choice, or anthropogenic intervention, had a significant impact on the size of the avalanches in counterintuitive ways. When the agent's intention is to create a large avalanche, the average avalanche size is decreased relative to the random deposition. On the other hand, if the agent's intention is to cause the smallest possible avalanche, the result is a major increase in the chances of triggering exceptionally large avalanches. The relevance between the sandpile model and forest fires is that the buildup of fuel functions like the buildup of sand, releasing in critical events. The agent's intention effect in the sandpile experiment is in agreement with the observation in anthropogenic wildfire management, where aggressively suppressing small early season fires can facilitate conditions that significantly increase the rate of large and destructive events (Yoder *et al.* 2011). To expand the conceptual model, the setting of prescribed fires can be considered analogous to an agent's intention to cause a collapse of the sandpile. Here, for the first time, we leverage this powerful model to quantify the effect of prescribed fire, a major intervention of forest management, on the incidence of wildfires. A series of simulations and a comprehensive analysis of historical data of forest fires from multiple states demonstrate that prescribed burns can significantly reduce the incidence of especially destructive wildfires.

Materials and methods

*Simulation*

<u>Basic sandpile model</u>

The basic sandpile model for describing natural wildfires without anthropogenic effects is defined as follows:

(i) A sand grain drops onto a random position on a $L$ x $L$ square grid, for example, a 20 x 20 grid.

(ii) Each position can hold up to 3 sand grains. If a position has more than 3 sand grains, the sandpile will collapse and these sand grains will be evenly distributed to cardinally adjacent sites.

(iii) A collapse of a sand pile at one site can cause collapse at adjacent sites, leading to a propagation across the plane until the grid reaches equilibrium.

(iv) A border hit, referred to as an "avalanche" or "event of interest," is when a propagating collapse causes a sand pile at any edge site to collapse. The number of border hits is recorded as the size of the event. Any sand grains which would fall off the edge are lost.

(v) Simulation process: repeat steps (i) through (iv) for a large number of times, for example, 64,000 sequentially dropped sand grains onto the grid.

(vi) The observed event sizes and their corresponding frequencies during the simulation process are summarized for further analysis.

The random drops of sand grains represent growth of the trees in the forest, whereas local collapses may be regarded as small-scale natural wildfires in the areas where too much fuel has been piled up. When the tension due to fuel accumulation keeps being built up, a local collapse (wildfire) may be propagated very quickly, yielding an avalanche (destructive wildfire). A simple example of implementing the basic sandpile model (dropping a single sand grain) on a 5 x 5 square grid is depicted in Fig. 1. In this case, dropping a sand grain to the center position of a loaded grid causes a series of collapses, which eventually led to an event of size 1.

Sandpile model with intervention

We incorporated the periodic prescribed fires (anthropogenic wildfire intervention) to the basic sandpile model by simply emptying the positions with 3 sand grains (risky pileup) at different time intervals, i.e., between dropping every 2, 4,5 8, 10, 16, 20, 25, 50, 100, 200, 400, 800, 1600, 3200, 6400, and 12800 sand grains, respectively, as well as a system with no interventions or no burning (NB). These numbers were chosen to make the total number of sand grains, 64,000, divisible by each. This setting in the simulation can be considered analogous to a planned prescribed burn in a region with various frequencies, from high to low, respectively.

*Historical fire data*

We used fire data from Florida, California, and Georgia. Wildfire data from Florida was available from the Florida Forest Service (http://fireinfo.fdacs.gov/fmis.publicreports), with additional data provided by the Prescribed Fire Manager of the Florida Forest Service, John Saddler (personal communication, December 18, 2020). Wildfire data from California was available from CalFire (https://www.fire.ca.gov/stats-events/) as was prescribed fire data (https://gis.data.cnra.ca.gov/). Wildfire data from Georgia was provided by Fire Chief Frank Sorrells of the Georgia Forestry Commission (personal communication October 21, 2020). The fire incidence data from each state was divided for each year into seven class sizes, A-G, as defined by the federal government. We did not use Class A fires (<0.25 acres [<0.001 km$^2$]) in the analysis as they are likely heavily underreported (Ricotta *et al.* 1999) and so the numbers associated with this class in the data severely underestimates their true incidence. The maximum fire size in each class was used to represent that class for the subsequent analysis. Classes B, C, D, E, and F were considered to be 9.9, 99, 299, 999 and 4,999 acres, respectively (0.04. 0.40, 1.21, 4.04. and 20.2 km$^2$, respectively). Class G, which included wildfires at least 5,000 acres (>20.2 km$^2$), was set to be 10,000 acres (40.47 km$^2$) for ease of analysis. Prescribed and wildfire data was available for Florida from 1993 to 2019, for California from 1963 to 2019, and for Georgia from 1995 to 2020, all inclusive. Additional wildfire data, but no prescribed fire data was available for Florida from 1981 to 1992, inclusive. Complete data can be found in Table S1 (Supplementary Materials).

*Log$_{10}$ transformation based on sandpile model of self-organized criticality*

For each state (Florida, California, and Georgia), we divided yearly data into quintiles based on the number of acres subject to prescribed burning that year. The breakdown of the data from all three states is shown in Table 1. Based on the theorem described in the sandpile model of self-organized criticality by Bak *et al*. (1987), we took the log$_{10}$ of the average number of fires in each class for each category and plotted them against the log$_{10}$ of the maximum fire size in acres in that class for that category. For each plot we calculated the slope of the fitted linear regression line as determined by the sum of least squares, which represents the relative risk for destructive fires. The estimated values and standard errors of these slopes were analyzed for model comparisons and for hypothesis testing.

*Comparison of data with and without prescribed fires*

The Florida data consists of data without the record of prescribed fires (1981-1992) and data with the record of prescribed fires (1993-2019). To compare subsets of fire incidence data in Florida, we split the data with the record of prescribed fires into first (1993-2005) and second (2006-2019) halves, yielding three consecutive periods for comparisons, i.e., period I (1981-1992), period II (1993-2005), and period III (2006-2019). We then plotted the data from each subset in a box plot for visual examination (Fig. 3).

We used a two-sample *t* test of analysis of variance (ANOVA) to compare the Florida data with and without the record of prescribed fires. The nominal $p < 0.05$ was used to claim a significant difference between any comparison.

*Analysis of slopes*

We conducted the pairwise comparison of slopes of the fitted regression lines for various categories using the pooled *t*-test. Suppose we would like to test whether two slopes are identical, i.e., $H_o: b_1 = b_2$ vs. $H_a: b_1 \neq b_2$, where $b_1$ and $b_2$ are slopes estimated from two linear regression models, respectively. Under the null hypothesis, the *t*-test statistic

$$t = \frac{b_1 - b_2}{\sqrt{s_{b_1}^2 - s_{b_2}^2}} \qquad (2)$$

follows a *t* distribution with degree of freedom $n_1 + n_2 - 4$, where $s_{b_1}$ and $s_{b_2}$ represent the standard errors for two slopes and $n_1$ and $n_2$ are the sample sizes for two models, respectively.

The R script, named 'TwoSlope.ttest', for implementing the comparison of two slopes of fitted regression lines is available in Supplementary File S2.

We used simple linear regression to analyze the association between the estimated slopes (risk of destructive wildfire) and the average acres per year that were subject to prescribed burn in various categories. The significant negative slope of the regression plots indicates that increased prescribed burns will decrease the risk of destructive fires.

Results

*Simulated study*

Following the simulation strategy described in Materials and Methods, we randomly dropped a total of 64,000 sand grains onto a 20 x 20 grid in each simulation scenario and monitored the fire sizes (number of border hits) and their frequencies (numbers of occurrence) for subsequent analysis and comparison. Graphs showing the relationship between fire sizes and the logarithm of their frequencies based on a single simulation when prescribed burns (or interventions) were administered at various intervals can be found in Fig. S1 (Supplementary Material). This simulation scheme has been repeated three times and the mean and standard deviation of slopes for each scenario are summarized in Fig 2. When prescribed fires were sparsely administered, i.e., for every 12800, 6400, 3200, 1600, or 800 sand drops, the estimated values for their slopes did not differ from the slope when no prescribed fire was applied (NB). While passing a threshold of every 800 sand drops, the slope values decreased in an approximately linear fashion as the frequency of prescribed fires increased. This indicated that the chance of destructive wildfires can be substantially reduced if prescribed burns are administered more often. When prescribed fires were used more often than every 50 sand drops, there were not consistently at least three differently sized events, so slopes could not be confidently determined, and the slopes for those data points were not used. Only when prescribed burns were administered every 200 sand drops or more sparsely did events of at least size 5 (at least five sand grains hitting the borders) occur consistently.

*Comparison of data with and without record of prescribed fires*

The Florida data consists of data without the record of prescribed fires (period I, 1981-1992) and data with the record of prescribed fires (1993-2019). We split the data with the record of prescribed fires into first (period II, 1993-2005) and second (period III, 2006-2019) halves, yielding three consecutive periods for comparison. Fig. 3*a* indicates that there was a significant increase in the total size of prescribed burns between period II (1998-2008) and period III (2009-2019). We assume that the undocumented prescribed fires in period I (1981-1993) were significantly less than periods II or III. The comparison of the total areas burnt yearly between these three periods are shown by boxplot in Fig. 3*b*. Although no significant difference in means was detected among these three periods, it is clear to see a trend in the reduction of giant wildfires as the prescribed burns have been progressively introduced to forest management.

*Analysis of slopes calculated from sandpile model*

For the $\log_{10}$ transformed Florida data, the scatter plots with the best fitted linear regression lines for the five quintiles are shown in Fig. 4*a-e*, with the slope of slopes shown in Fig. 4*f*. The intersects, estimated slopes, standard errors, and p-values are shown in Table 2. We carried out a pairwise comparison among slopes of the fitted regression lines for various categories using the $\log_{10}$ transformed Florida data, the results of which are shown in Table 3. Due to the limited sample size (n = 6) in each category, only marginally significant differences (0.05 < p value < 0.1) have been detected between quintile 1 with quintile 2 and between quintile 1 with quintile 4. Similarly, due to the limited sample size, the regression of these slopes on the median land areas that were subject to prescribed burn in five categories only showed a trend of negative association (slope of -0.17 with p = 0.298 shown in Fig. 4), suggesting that increased prescribed burning tended to reduce the risk of destructive wildfires. Note that the magnitude of the slope for quintile 2, colored red in Fig. 4, appeared to be much lower than expected. Possible explanations for this outlier slope associated with the data in quintile 2 are proposed in the Discussion.

For the $\log_{10}$ transformed California data, the slope of slopes for the five quintiles is shown in Fig. 4*g*. The intersects, the slopes estimated for these five categories, their standard errors, and p-values are shown in Table 4. We did not carry out a pairwise comparison among slopes for California data since these slope values were close to one another with a slope of slopes that was nearly flat. The slope of the line of best fit in all quintiles ranged from -0.84 to -0.75, with an average p-value of 0.00023. Similar results were obtained when investigating the Georgia data, the slope of slopes of which is shown in Fig. 4*h*, where the slopes of the line of best fit in all quintiles ranged from -1.66 to -1.29 with an average p-value of 0.0066. The intersects, the slopes estimated for these five categories, their standard errors, and p-values are shown in Table 5. The differences between the results from analyzing Florida data and the results from analyzing California data or Georgia data are likely due to greatly varied overall size of prescribed burns in these states, as indicated in Table 6. Scatter plots with the best fitted linear regression lines for the five categories in both California and Georgia can be found in Figs. S2 and S3 (Supplementary Materials).

Discussion

The analysis of our simulated data (Figs. 2 and S2) as well as the historical data, including Florida data (Figs. 3 and Fig 4*a-f*), California data (Figs. 4*g* and S3) and Georgia data (Figs. 4*h* and S3), supports the previous claim that the behavior of natural wildfires can be well described using the sandpile model (Ricotta *et al.* 1999; Malamud 2004; Yoder *et al.* 2011). For the first time, we used both simulated data (Fig. 2) and historical forest wildfire data (Fig. 4) to demonstrate that the sandpile model may be also used to test whether prescribed fires reduce the risk of destructive wildfires. These results suggested that the application of this model may be expanded to investigate the influence of other wildfire drivers, such as temperature, rainfall, atmospheric circulation patterns, and socioeconomic factors (Pereira *et al*. 2005; Costa *et al*. 2011). To facilitate such research, we have developed an R program, which researchers can use to study the association between wildfire outcome and any factor of interest using simulations. Outcomes of natural wildfires are often the results of influences from multiple factors. For this reason, the current version of the simulation program is not powered enough to handle the complexity due to the potential interactions between factors, for example, temperature and rainfall in a specific landform. Since prescribed burns are an anthropogenic intervention rather than a natural condition, it may be regarded as a factor independent of other drivers. Therefore, the current simulation program is suitable for studying the relatedness between prescribed fires and outcomes of wildfires. In future studies, we intend to develop advanced simulation schemes to extend the sandpile model for analyzing data with interacting factors for wildfires or other complications.

Previous studies applied mathematical models to temporal and spatial data of local fires (microenvironment) and identified a strong association between past and current fires, indicating that recently burned areas may be less flammable than those not recently burned (Malamud *et al.* 2005; Bradstock *et al.* 2012; Price *et al.* 2012; Price *et al.* 2015). However, it is challenging to expand this strategy to estimate the general effect of recent fires on future fires at a much larger scale, such as statewide or nationwide levels, for two reasons: (1) various regions are highly heterogeneous in many aspects, including landscape, climate, fire history, and anthropogenic effects; (2) existing wildfire database statistics are often low in spatial and/or temporal resolution of their data sets. Therefore, a causal model that may be overfitted to local data likely has limited predictive ability when applied to other regions with different microfeatures.

In our study, we adapted the sandpile model to study wildfires at statewide scale, in which the connection between past and current fires and the gradual buildup of fuel between are well represented by local collapses of sand piles, redistribution of the collapsed sand grains, and rebuild of sand piles at these loci. One major advantage of using the power-law represented by the sandpile model to describe natural hazards, including earthquakes, asteroid impacts, volcanic eruptions and wildfire (the subject of this manuscript), is leveraging a simple model to account for complex systems which are determined by many interacting factors. Note the time of fuel build-up varies geographically and is likely to be different in, for example, Florida and California, reflecting their respective natural resistances to wildfire. Our simulation suggests that such baseline features of Florida and California may be reflected by the slopes of fire categories when no prescribed fires (NB) were applied in the two states.

The simulation shows a correlation between an increase in prescribed burning and a decrease in the slope value, i.e., a more negative slope (Fig 2), given that prescribed fires are administered within a certain range. Only when prescribed burns were administered every 200 sand drops less often did events of at least size 5 (at least five sand grains hitting the borders) occur regularly. This result indicates that (1) insufficient intervention fires are not effective in reducing the risk of gigantic wildfires and, on the other hand, (2) excessive burns cannot further increase such an effect, resulting in wasted cost and effort. These conclusions are supported by the analysis of wildfire data in multiple states, including Florida, California and Georgia. The slope of the Florida first quintile is -0.81, the second -1.01, the third -0.92, the fourth -0.97, and the fifth -0.94. The negative slope of slopes observed in Florida data demonstrates that increasing the number of acres subject to prescribed burning has substantially reduced the incidence of destructive wildfires. This tracks with the increasing use of prescribed burning in Florida over the past twenty-seven years, while use has remained essentially static in California and Georgia at low and high levels, respectively. From 1993 to 2002, an average of about 1.9 million acres (7689 km$^2$) in Florida were subject to prescribed burning, while from 2010 to 2019 an average of about 2.3 million acres (9308 km$^2$) were subject, an increase of ~20%. Compared to Florida, the intervention fires appeared to be far from enough in California to be effective in preventing gigantic wildfires. As shown in Table 6, the median prescribed burn in California was dramatically less than that in Florida. While the Georgia prescribed fire data for the fifth quintile was comparable to the first quintile in Florida. It may be that California is in the "low burn" section of Fig. 2, while Georgia is in the "high burn" section. These statistics of historical prescribed fires may explain why, unlike the result from Florida data, the slopes for five quintiles were invariant when California data or Georgia data were analyzed.

Interestingly, the slope of the log$_{10}$ of fire incidence of the second quintile in Florida is far lower than would be expected based on the trend present in the other data (Fig. 4*f*). The slope of the second quintile is -1.01, lower even than the fifth quintile, which had a slope of -0.94. It may be that the major dip indicates the presence of an optimal level of prescribed burning. The possible presence of optima requires further study to determine if judicious application of prescribed burning can be more effective, both in terms of cost and effort, than generally increasing the use of prescribed burning. Our data and conclusions warrant future research involving the use of this model to develop the optimal strategy, with both geographic and temporal considerations, for prescribing fires which can achieve the maximum efficacy of suppressing giant wildfires with minimum possible human efforts and operative costs.

In all examined fire incidence data, Class A fires (<0.25 acres [0.001 km$^2$]) were greatly underrepresented relative to expectations based on the modeling. This is likely because many, if not most, small fires are not reported to statewide agencies and may be put out by landowners or simply run out of fuel and sputter out without intervention. The national fire policy in the United States from the late 1800's has been fire suppression, leading to significant fire deficits in the 20$^{th}$ century. This fire suppression policy has been demonstrably linked to major increases in fire intensity, especially in the 21$^{st}$ century (Haugo *et al*. 2019; Roos *et al*. 2020). The increased incidence of especially destructive fires could be due to buildup of debris on the forest floor. It should be noted that not all the blame for increased fire intensity is due to fire suppression; indeed, global climate change has greatly stressed forests by shifting rainfall patterns, causing prolonged droughts (Stephens *et al*. 2018) and opportunistic infestations, such as by bark beetles (Preisler *et al*. 2017), leading to large stands of dead trees, which fuel the more intense fires. The fire regime in California can be compared to the fire regime in Florida, which has made aggressive use of prescribed burns. The results show that not only does the fire incidence track the expected negative linear $\log_{10}$ line, but that in years in which more acreage was subject to prescribed burns the slope of the line of best fit of the $\log_{10}$ line is more negative, indicating across the board reductions in larger and more destructive fires.

Data Availability

Complete wildfire and prescribed fire data can be found in Supplementary File S1. The R script, named 'TwoSlope.ttest', for implementing the comparison of two slopes of fitted regression lines is available in Supplementary File S2. The R script, named 'Sandpile_wildfire', for implementing the simulations of wildfire model with an impacting factor is available in Supplementary File S3.

Acknowledgements

We would like to thank Dr. Zhenyu Jia at University of California Riverside for his insightful comments and suggestions during our discussion on this project, John Saddler of the Florida Forest Service and Frank Sorrells of the Georgia Forestry Commission for providing fire data, and John H. Miller for bringing the phenomenon of self-organized criticality to our attention. This research did not receive any specific funding.

Conflicts of Interest

The authors declare no conflicts of interest.

Table 1. The average number of wildfires of each class across subgroups in Florida and California defined based on the acreage of prescribed burns. 1 acre = 0.00404 km².

| Class | Class Size | Category based on acres of prescribed burn (Median of category in acre) | | | | |
|---|---|---|---|---|---|---|
| | | Quintile 1 | Quintile 2 | Quintile 3 | Quintile 4 | Quintile 5 |
| *Florida* | | | | | | |
| | | (1,779,524) | (2,055,513) | (2,130,843) | (2,294,177) | (2,475,650) |
| A | 0.24 | 1145.17 | 680.00 | 986.60 | 764.80 | 534.00 |
| B | 9.9 | 2572.67 | 1621.60 | 2381.20 | 1964.40 | 1392.50 |
| C | 99 | 795.83 | 449.60 | 734.40 | 592.80 | 357.50 |
| D | 299 | 109.50 | 46.80 | 97.20 | 73.80 | 36.83 |
| E | 999 | 61.83 | 20.40 | 41.60 | 30.60 | 17.33 |
| F | 4999 | 23.17 | 7.20 | 13.80 | 10.20 | 5.33 |
| G | 10000 | 10.50 | 1.20 | 4.40 | 2.20 | 2.33 |
| | | | | | | |
| *California* | | | | | | |
| | | (2,844) | (10,095) | (21,694) | (32,402) | (47,786) |
| A | 0.24 | 5908.5 | 3928.5 | 4190.9 | 4272.4 | 2697.9 |
| B | 9.9 | 1763.8 | 1500.1 | 1425.7 | 1505.1 | 1486.6 |
| C | 99 | 373.2 | 300.2 | 261.4 | 256.6 | 232.0 |
| D | 299 | 74.1 | 66.2 | 51.6 | 47.2 | 47.5 |
| E | 999 | 36.7 | 31.9 | 24.9 | 24.3 | 26.4 |
| F | 4999 | 15.3 | 13.0 | 13.0 | 10.9 | 15.8 |
| G | 10000 | 5.6 | 5.0 | 4.4 | 4.0 | 6.5 |
| | | | | | | |
| *Georgia* | | | | | | |
| | | (1,023,396) | (1,252,156) | (1,351,109) | (1,469,347) | (1,632,603) |
| A | 0.24 | 1451.33 | 2151 | 1780.2 | 1845.6 | 1743.4 |
| B | 9.9 | 3149.33 | 4792.2 | 4177 | 3947.2 | 4242.8 |
| C | 99 | 372.17 | 563.2 | 568.4 | 487.6 | 583 |
| D | 299 | 12.33 | 18.8 | 28.8 | 20 | 24.2 |
| E | 999 | 3.5 | 2.8 | 6 | 4 | 4.4 |
| F | 4999 | 1.17 | 0.001 | 2.6 | 0.6 | 2 |
| G | 10000 | 0.33 | 0.001 | 0.6 | 0.001 | 0.8 |

Table 2. Estimation of slopes and relevant metrics for quintiles of Florida wildfire data.

| Category | Estimate of Slope | Standard Error | P value | $R^2$ |
|---|---|---|---|---|
| Quintile 1 | -0.8060 | 0.0714 | 0.00035 | 0.9696 |
| Quintile 2 | -1.0139 | 0.1034 | 0.00061 | 0.9601 |
| Quintile 3 | -0.9152 | 0.0769 | 0.00029 | 0.9726 |
| Quintile 4 | -0.9687 | 0.0904 | 0.00043 | 0.9663 |
| Quintile 5 | -0.9448 | 0.0803 | 0.00030 | 0.9719 |

Table 3. The pairwise comparison of slopes among five categories in Florida data.
The diagonal entries (shaded cells) are the estimated slopes; the upper triangle show the p values for the pairwise comparisons; the '*' and '-' in the lower triangle represent significant results and insignificant results, respectively based on p value < 0.1.

| Category | Quintile 1 | Quintile 2 | Quintile 3 | Quintile 4 | Quintile 5 |
|---|---|---|---|---|---|
| Quintile 1 | -0.8060 | *0.0683* | *0.1641* | *0.0978* | *0.1163* |
| Quintile 2 | * | -1.0139 | *0.2329* | *0.3754* | *0.3061* |
| Quintile 3 | - | - | -0.9152 | *0.3321* | *0.3984* |
| Quintile 4 | * | - | - | -0.9687 | *0.4241* |
| Quintile 5 | - | - | - | - | -0.9448 |

Table 4. Estimation of slopes and relevant metrics for quintiles of California wildfire data.

| Category | Estimate of Slope | Standard Error | P value | $R^2$ |
|---|---|---|---|---|
| Quintile 1 | -0.8164 | 0.0547 | 0.00012 | 0.9824 |
| Quintile 2 | -0.8091 | 0.0484 | 0.00008 | 0.9859 |
| Quintile 3 | -0.8091 | 0.0643 | 0.00023 | 0.9754 |
| Quintile 4 | -0.8352 | 0.0627 | 0.00018 | 0.9780 |
| Quintile 5 | -0.7562 | 0.0742 | 0.00052 | 0.9629 |

Table 5. Estimation of slopes and relevant metrics for quintiles of Georgia wildfire data.

| Category | Estimate of Slope | Standard Error | P value | $R^2$ |
|---|---|---|---|---|
| Quintile 1 | -1.3456 | 0.1369 | 0.0006 | 0.9603 |
| Quintile 2 | -1.6576 | 0.2869 | 0.0287 | 0.9435 |
| Quintile 3 | -1.2908 | 0.1244 | 0.0005 | 0.9642 |
| Quintile 4 | -1.4918 | 0.1544 | 0.0024 | 0.9689 |
| Quintile 5 | -1.2899 | 0.1429 | 0.0008 | 0.9532 |

Table 6. Statistics of prescribed (Rx) burns in California, Florida and Georgia.

| State | State Area (acres) | Slope of slopes | Median Rx burns (acres) | | % total area | |
|---|---|---|---|---|---|---|
| | | | Quintile 1 | Quintile 5 | Quintile 1 | Quintile 5 |
| CA | 104,765,269 | 0.9481 | 2,844 | 47,786 | 0.0027 | 0.0456 |
| FL | 42,085,120 | -0.1748 | 1,779,524 | 2,475,650 | 4.23 | 5.88 |
| GA | 38,032,000 | 0.1542 | 1,023,396 | 1,632,603 | 2.69 | 4.29 |

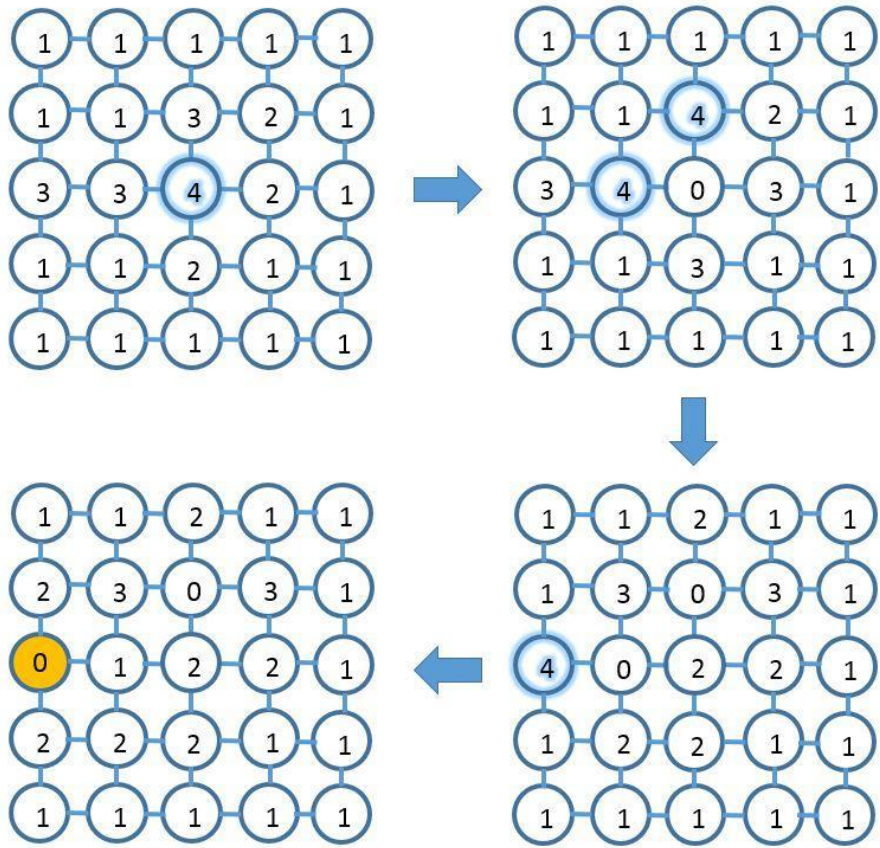

**Fig. 1.** An example of implementing the basic sandpile model (drop a single sand grain) on a 5 x 5 square grid. The number at each position represent the number of sand grains. Positions with glowing blue represent the unstable positions subject to the redistribution of sand grains collected in these positions. A border hit (event of interest) is denoted by an orange position along the edge of the grid, and the number of border hits represents the event size, 1 in this example.

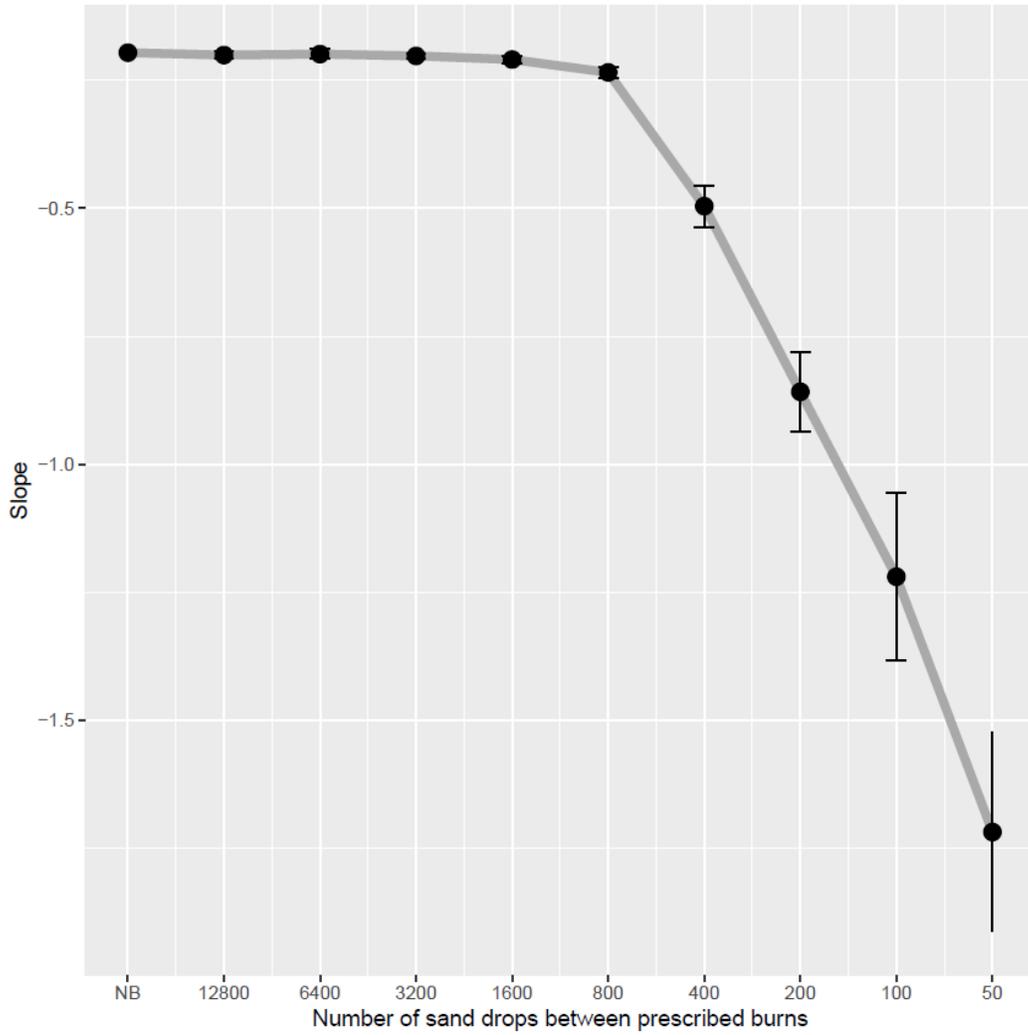

**Fig. 2.** Slopes of each tested intervention frequency on a 20x20 grid. Error bars indicate standard deviation. Interventions occurring more often than every 50 sand drops are excluded as those did not consistently have at least three different data points to confidently assign slopes.

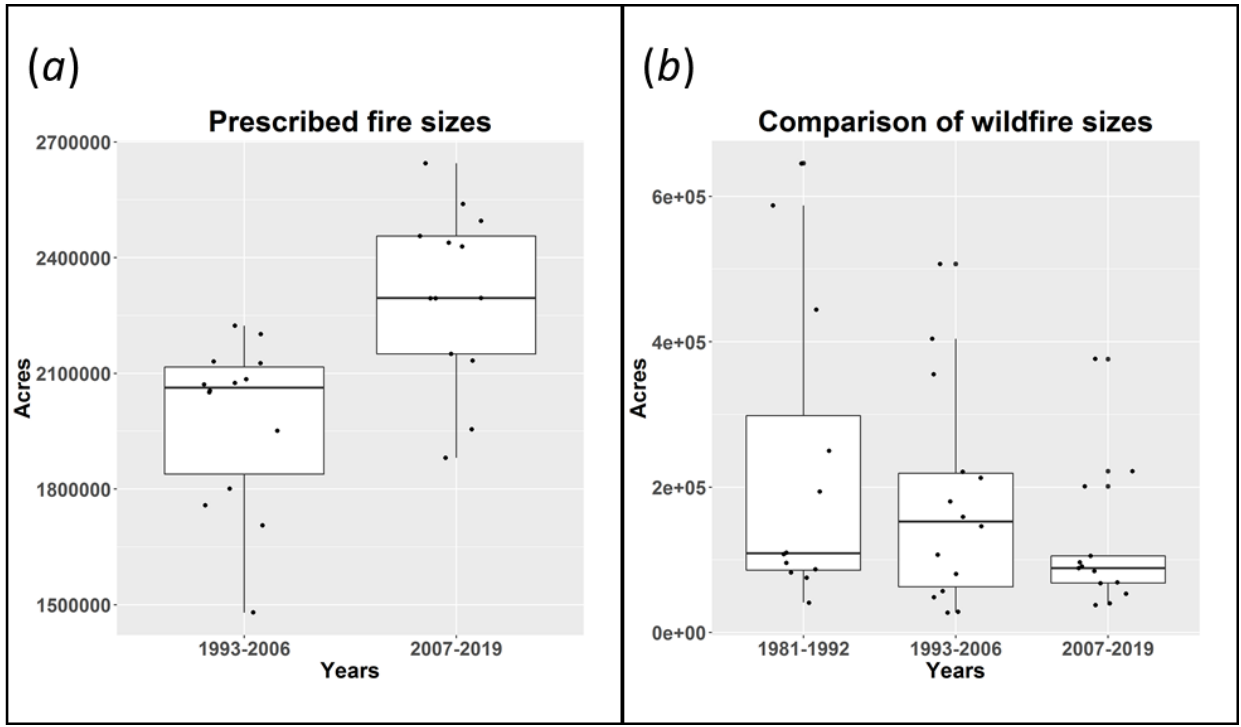

**Fig. 3.** Comparison of Florida wildfire data between various periods. (*a*) The areas of prescribed fires in period III (2006-2019) were significantly greater than those in period II (1993-2005), with p value = 0.0007. (*b*) No significant difference in means were detected between three periods (p values > 0.05). There is a trend in the reduction of destructive wildfires (extreme values along y axis) along three consecutive periods. 1 acre = 0.00404 km$^2$.

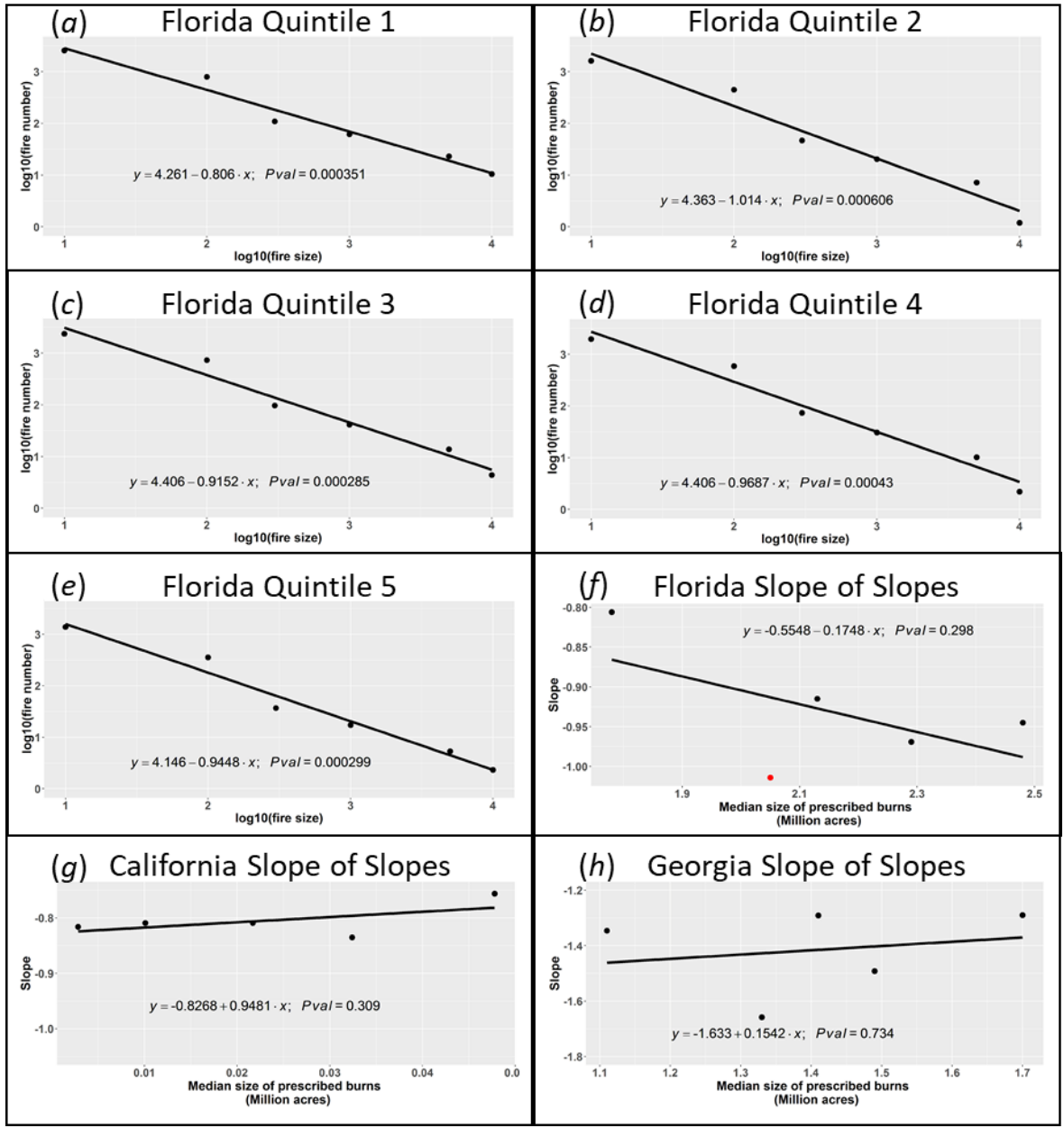

**Figure 4**. Analysis of slopes based on wildfire data. (*a-e*) Fitted regression models for the Florida wildfire data in quintiles, based on the number of acres subject to prescribed burning. (*f*) Fitted regression model for the Florida slopes vs. the median sizes of prescribed burns, with data point for quintile 2 being labelled with red. (*g*) Fitted regression model for the California slopes vs. the median sizes of prescribed burns. (*h*) Fitted regression model for the Florida slopes vs. the median sizes of prescribed burns